\documentclass{Interspeech2024}

\usepackage{amsmath,graphicx,amsfonts,amsthm,amssymb,stfloats,booktabs,multirow,bbding,pifont,url,color,makecell,bm}

\newcommand{\cmark}{\ding{51}}%
\newcommand{\xmark}{\ding{55}}%

\usepackage{cite} 
\usepackage{multirow}
\usepackage{algorithm}
\usepackage{algpseudocode}
\usepackage{colortbl}  
\usepackage{tabularx}
\usepackage[table,xcdraw]{xcolor}
\usepackage{subcaption}
\usepackage[normalem]{ulem}

\interspeechcameraready

\title{Noise-aware Speech Enhancement using Diffusion Probabilistic Model}

\name{Yuchen Hu$^1$, Chen Chen$^1$, Ruizhe Li$^2$, Qiushi Zhu$^3$, Eng Siong Chng$^1$}

\address{
  $^1$Nanyang Technological University, Singapore \quad $^2$University of Aberdeen, UK \\
  $^3$University of Science and Technology of China, China}
\email{yuchen005@e.ntu.edu.sg}


\begin{document}

\maketitle

\begin{abstract}
With recent advances of diffusion model, generative speech enhancement (SE) has attracted a surge of research interest due to its great potential for unseen testing noises. However, existing efforts mainly focus on inherent properties of clean speech, underexploiting the varying noise information in real world. In this paper, we propose a noise-aware speech enhancement (NASE) approach that extracts noise-specific information to guide the reverse process in diffusion model. Specifically, we design a noise classification (NC) model to produce acoustic embedding as a noise conditioner to guide the reverse denoising process. Meanwhile, a multi-task learning scheme is devised to jointly optimize SE and NC tasks to enhance the noise specificity of conditioner. NASE is shown to be a plug-and-play module that can be generalized to any diffusion SE models. Experiments on VB-DEMAND dataset show that NASE effectively improves multiple mainstream diffusion SE models, especially on unseen noises\footnote{Code, model and processed audio files are publicly available at: https://github.com/YUCHEN005/NASE}.
\end{abstract}

\section{Introduction}
\label{sec:intro}
Speech enhancement (SE) aims to estimate clean speech signals from audio recordings that are corrupted by acoustic noises~\cite{hendriks2013dft}, which usually serves as a front-end processor in many real-world applications, including speech recognition~\cite{hu2022interactive,hu2023gradient}, hearing aids~\cite{fedorov2020tinylstms} and speaker recognition~\cite{hansen2015speaker}.
With advances of deep learning in the past decade, significant progress has been made.

Deep learning based SE can be roughly divided into two categories, based on the criteria used to estimate the transformation from noisy speech to clean speech.
The first category trains discriminative models to minimize the distance between noisy and clean speech.
However, as supervised methods are inevitably trained on a finite set of training data with limited model capacity for practical reasons, they may not generalize well to unseen situations, \textit{e.g.}, different noise types, different signal-to-noise ratios (SNR) and reverberations.
Additionally, some discriminative methods have been shown to result in undesirable speech distortions~\cite{wang2019bridging}.

The second category trains generative models to learn the distribution of clean speech as a prior for speech enhancement, instead of learning a direct noisy-to-clean mapping.
Several approaches have employed deep generative models for speech enhancement, including generative adversarial network (GAN)~\cite{pascual2017segan,baby2019sergan}, variational autoencoder (VAE)~\cite{fang2021variational,bie2022unsupervised} and flow-based models~\cite{nugraha2020flow,strauss2021flow}.
Recent advances of diffusion probabilistic model have launched a new surge of research interest in generative SE~\cite{lu2021study,lu2022conditional,welker22speech,lemercier2022storm,richter2023speech,shi2023diffusion}.
The main principle of these approaches is to learn the inherent properties of clean speech, such as its temporal and spectral structure, which then serve as prior knowledge to infer clean speech from noisy input.
Therefore, they focus on generating clean speech and are thus considered more robust to varying acoustic conditions in the real world.
Existing studies~\cite{fang2021variational,bie2022unsupervised,lu2022conditional} have showed better performance of generative SE on unseen testing noises than discriminative counterparts.
However, these approaches fail to fully exploit the noise information inside input noisy speech~\cite{zhang2023noise}, which could be instructive to the denoising process of SE, especially in unseen test conditions.

In this paper, we propose a noise-aware speech enhancement (NASE) approach that extracts noise-specific information to guide the reverse process of diffusion model.
Specifically, we design a noise classification (NC) model and extract its acoustic embedding as a noise conditioner for guiding the reverse denoising process~\cite{dhariwal2021diffusion}.
With such noise-specific information, the diffusion model can target at the noise component in noisy input and thus remove it more effectively.
Meanwhile, a multi-task learning scheme is devised to jointly optimize SE and NC tasks, which aims to enhance the noise specificity of extracted noise conditioner.
Our NASE is shown to be a plug-and-play module that can be generalized to any diffusion SE models for improvement.
Experiments verify its effectiveness on multiple diffusion backbones, especially on unseen testing noises.

\section{Diffusion Probabilistic Model}
\label{sec:diffusion}
In this section, we briefly introduce the vanilla diffusion probabilistic model in terms of its \emph{diffusion} and \emph{reverse} processes.
To formulate speech enhancement task, we define the input noisy speech as $y$ and its corresponding clean speech as $x_0$. 
Generally speaking, SE aims to learn a transformation $f$ that converts the noisy input to clean signal: $x_0 = f(y),\ \{x_0, y\} \in \mathbb{R}^L$, $L$ is signal length in samples. 

\noindent\textbf{Diffusion process} is defined as a $T$-step Markov chain that gradually adds Gaussian noise to original clean signal $x_0$:
\vspace{-0.1cm}
\begin{equation}
  q(x_1, \cdots, x_{T}|x_0) = \prod_{t=1}^{T} q(x_t|x_{t-1}),
  \label{eq1}
  \vspace{-0.1cm}
\end{equation}

\noindent with a Gaussian model $q(x_t|x_{t-1}) = \mathcal{N}(x_t;$ $\sqrt{1-\beta_{t}}x_{t-1},\beta_{t}I)$, where $\beta_{t}$ is a small positive constant serving as a pre-defined schedule. 
After sufficient diffusion steps $T$, the clean $x_0$ is finally converted to a latent variable $x_T$ with an isotropic Gaussian distribution $p_\text{latent}(x_T)=\mathcal{N}(0,I)$. 
Therefore, conditioned on $x_0$, the sampling distribution of step $t$ in the Markov chain can be derived as:
\begin{equation}
  q(x_t|x_0) = \mathcal{N}(x_t;\sqrt{\bar{\alpha}_t}x_0,(1-\bar{\alpha}_t)I),
  \label{eq2}
\end{equation}
where $\alpha_t = 1-\beta_t$ and $\bar{\alpha}_t = \prod_{s=1}^{t} \alpha_s$. 

\noindent\textbf{Reverse process} aims to restore $x_0$ from the latent variable $x_T$ based on the following Markov chain:
\vspace{-0.2cm}
\begin{equation}
  p_\theta(x_0, \cdots, x_{T-1}|x_T) = \prod_{t=1}^{T} p_{\theta}(x_{t-1}|x_t),
  \label{eq3}
\end{equation}

\noindent where $p_\theta(\cdot)$ is a distribution of reverse process with learnable parameters $\theta$.
As marginal likelihood $p_{\theta}(x_0) = \int p_\theta(x_0, \cdots, x_{T-1}|x_T)\cdot p_\text{latent}(x_T) dx_{1:T}$ is intractable for calculation, the ELBO~\cite{ho2020denoising} is employed to approximate the objective function for model training.
Consequently, the equation of reverse process can be formulated as:
\begin{equation}
\begin{split}
\begin{aligned}
 p_{\theta}(x_{t-1}|x_t) &= \mathcal{N}(x_{t-1};\mu_\theta(x_t,t),\tilde{\beta}_t I), \\
 \text{where} \quad \mu_\theta(x_t,t) &= \frac{1}{\sqrt{\alpha_t}}(x_t-\frac{\beta_t}{\sqrt{1-\bar{\alpha}_t}}\epsilon_\theta(x_t,t))
 \label{eq4}
\end{aligned}
\end{split}
\end{equation}

\noindent The $\mu_{\theta}(x_t,t)$ predicts the mean of $x_{t-1}$ by removing the estimated Gaussian noise $\epsilon_{\theta}(x_t,t)$ in $x_t$, and the variance of $x_t$ is fixed to a constant $\tilde{\beta}_t = \frac{1-\bar{\alpha}_{t-1}}{1-\bar{\alpha}_t}\beta_t$.

\section{Methodology}
\label{sec:method}
\begin{figure*}[t]
\centering
\includegraphics[width=0.9\textwidth]{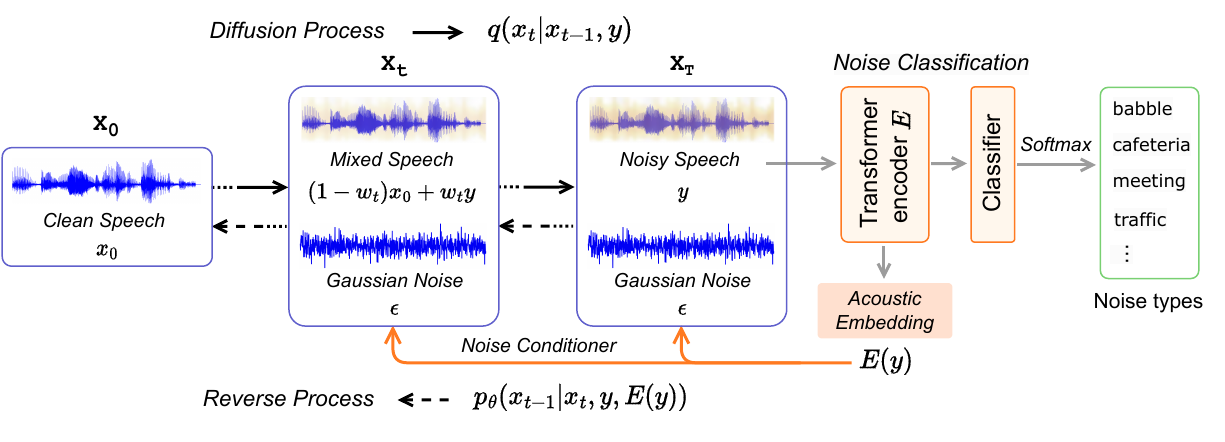}
\vspace{-0.4cm}
\caption{The overall framework of our proposed NASE approach.}
\label{fig1}
\vspace{-0.4cm}
\end{figure*}

In this section, we introduce our proposed NASE approach, which integrates a noise conditioner from classification module into the reverse process of conditional diffusion model for guidance. The overall framework of NASE is shown in Fig~\ref{fig1}.

\subsection{Conditional Diffusion Probabilistic Model}
\label{ssec:cond_diffusion}
Considering that real-world noises usually does not obey the Gaussian distribution, recent study~\cite{lu2022conditional} propose conditional diffusion probabilistic model that incorporates the noisy data $y$ into both diffusion and reverse processes.
Specifically, a dynamic weight $w_t\in[0,1]$ is employed for linear interpolation from $x_0$ to $x_t$.
As shown in Fig.~\ref{fig1}, each latent variable $x_t$ contains three terms: clean component $(1-w_t) x_0$, noisy component $w_t y$ and Gaussian Noise $\epsilon$. Therefore, the diffusion process in Eq.~(\ref{eq2}) can be rewritten as:  
\begin{equation}
\begin{split}
\begin{aligned}
 q(x_t|x_0,y) &= \mathcal{N}(x_t;(1-w_t)\sqrt{\bar{\alpha}_t}x_0 + w_t\sqrt{\bar{\alpha}_t}y ,\delta_tI), \\
 \text{where} \quad \delta_t &= (1-\bar{\alpha}_t)-w_t^2\bar{\alpha}_t
 \label{eq5}
\end{aligned}
\end{split}
\end{equation}

\noindent Here $w_t$ starts from $w_0 = 0$ and gradually increases to $w_T \approx 1$, turning the mean of $x_t$ from clean speech $x_0$ to noisy speech $y$.

Starting from $x_T$ with distribution $\mathcal{N}(x_T,\sqrt{\bar{\alpha}_T}y,\delta_TI)$, the conditional reverse process can be formulated from Eq.~(\ref{eq4}) as:
\begin{equation}
  p_\theta(x_{t-1}|x_t,y) = \mathcal{N}(x_{t-1};\mu_\theta(x_t,y,t),\tilde{\delta}_tI),
  \label{eq6}
\end{equation}

\noindent where $\mu_{\theta}(x_t,y,t)$ predicts the mean of $x_{t-1}$. 
In contrast to vanilla reverse process, here the neural model $\theta$ considers both $x_t$ and noisy speech $y$ for prediction. 
Thus, similar to Eq.~(\ref{eq4}), the mean $\mu_{\theta}$ is defined as a linear combination of $x_t$, $y$, and $\epsilon_{\theta}$:
\begin{equation}
  \mu_\theta(x_t,y,t) = c_{xt}x_{t} + c_{yt} y - c_{\epsilon t} \epsilon_\theta(x_t,y,t),
  \label{eq7}
\end{equation}

\noindent where the coefficients $ c_{xt}, c_{yt}, $ and $ c_{\epsilon t} $ are derived from the ELBO optimization criterion in~\cite{lu2022conditional}. 
Finally, the Gaussian noise $\epsilon$ and non-Gaussian noise $y-x_0$ are combined as ground-truth $C_t$ to supervise the predicted $\epsilon_\theta$ from neural model:
\begin{align}
    &C_t(x_0,y,\epsilon) = \frac{m_t\sqrt{\bar{\alpha}_t}}{\sqrt{1-\bar{\alpha}_t}}{(y-x_0)} + \frac{\sqrt{\delta_t}}{\sqrt{1-\bar{\alpha}_t}}\epsilon \label{eq8}\\ 
    &\mathcal{L}_\text{diff} = \parallel \epsilon_\theta(x_t,y,t) - C_t(x_0,y,\epsilon) \parallel_1,
    \label{eq9}
\end{align}

\subsection{Noise Conditioner from Classification Module}
\label{ssec:noise_cond}
Based on conditional diffusion probabilistic model, we propose to fully exploit the noise-specific information inside noisy speech $y$ to guide the reverse denoising process.
In particular, inspired by prior work~\cite{dhariwal2021diffusion}, we design a noise classification module to produce acoustic embedding as conditioner, which informs the diffusion model about what kind of noise to remove.

As shown in Fig.~\ref{fig1}, the noisy speech $y$ is sent into a Transformer Encoder $E$ as well as a linear classifier for noise type classification, where the output acoustic embedding $E(y)$ of encoder are extracted out as a noise conditioner.
Furthermore, in order to ease the training of noise classification module and extract better acoustic embedding with rich noise-specific information, we load an audio pre-trained model called BEATs from prior work~\cite{chen2022beats} for encoder $E$.
It is pre-trained on large-scale AudioSet~\cite{gemmeke2017audio} dataset and thus can provide rich prior knowledge of audio noise for classification.

After extracting out the acoustic embedding, we send it to guide the reverse process as an extra conditioner.
Specifically, the Eq.~(\ref{eq6}) and (\ref{eq7}) can be re-written as:
\begin{equation}
\begin{split}
\begin{aligned}
 p_\theta(x_{t-1}|x_t,y,E(y)) = \mathcal{N}(x_{t-1};\mu_\theta(x_t,y,t,E(y)),\tilde{\delta}_tI), \\
 \mu_\theta(x_t,y,t,E(y)) = c_{xt}x_{t} + c_{yt} y - c_{\epsilon t} \epsilon_\theta(x_t,y,t,E(y)),
 \label{eq10}
\end{aligned}
\end{split}
\end{equation}

\noindent The predicted noise $\epsilon_\theta$ is also conditioned on acoustic embedding $E(y)$, which contains noise-specific information and thus enables more effective denosing in reverse process.
Specifically, we select three techniques to inject the noise conditioner into $\epsilon_\theta$, \textit{i.e.}, addition, concatenation and cross-attention fusion with original inputs $x_t$ and $y$.
The $\epsilon_\theta$ in Eq.~(\ref{eq9}) should also be re-written accordingly.

\subsection{Multi-task Learning}
\label{ssec:multitask}
To further enhance the noise specificity of acoustic embedding $E(y)$, we perform noise classification as an auxiliary task:
\begin{equation}
\begin{split}
\begin{aligned}
 \mathcal{L}_\text{NC} &= \text{CrossEntropy}(\hat{C}, C) \\
 \text{where} \quad \hat{C} &= \text{Softmax}(P(E(y)))
 \label{eq11}
\end{aligned}
\end{split}
\end{equation}

\noindent Here the $\hat{C}$ and $C$ denote the predicted probability distribution and noise class label respectively.
$P$ is a linear classifier.

Multi-task learning scheme is employed to optimize SE and NC tasks simultaneously for better generalization:
\begin{equation}
 \mathcal{L} = \mathcal{L}_\text{diff} + \lambda_\text{NC} \cdot \mathcal{L}_\text{NC}
 \label{eq12}
\end{equation}

\noindent where $\lambda_\text{NC}$ is a weighting hyper-parameter to balance two tasks.

\section{Experiments and Results}
\label{sec:exp_result}

\subsection{Experimental Setup}
\label{ssec:exp_setup}

\noindent\textbf{Dataset.} 
We evaluate the proposed approach on public VoiceBank-DEMAND (VBD) dataset~\cite{valentini2016investigating}.
In particular, the training set contains 11,572 noisy utterances from 28 speakers in VoiceBank corpus~\cite{veaux2013voice}, which are recorded at a sampling rate of 16 kHz and mixed with 10 different noise types at SNR levels of 0, 5, 10, and 15 dB. 
The test set contains 824 noisy utterances from 2 speakers, which are mixed with 5 types of unseen noise at SNR levels of 2.5, 7.5, 12.5, and 17.5 dB.
For further evaluation on unseen noise, we also simulated noisy test data with three types of noise from prior work~\cite{lin2021unsupervised}, \textit{i.e.}, ``Helicopter'', ``Baby-cry'' and ``Crowd-party''.

\noindent\textbf{Configurations.} 
We select three various types of open-sourced diffusion SE models as our backbone, including one conditional diffusion model: CDiffuSE\footnote{https://github.com/neillu23/CDiffuSE}~\cite{lu2022conditional}, and two score-based diffusion models: StoRM\footnote{https://github.com/sp-uhh/storm}~\cite{lemercier2022storm} and SGMSE+\footnote{https://github.com/sp-uhh/sgmse}~\cite{richter2023speech}, and we follow their best configurations as our backbone.
The pre-trained BEATs\footnote{https://github.com/microsoft/unilm/tree/master/beats} model contains 12 Transformer~\cite{vaswani2017attention} encoder layers, with 12 attention heads and 768 embedding units.
The number of noise types for classification is set to 10, following the training data.
The weight $\lambda_\text{NC}$ is set to 0.3.

\noindent\textbf{Metric.} 
We use perceptual evaluation of speech quality (PESQ)~\cite{rix2001perceptual}, extended short-time objective intelligibility (ESTOI)~\cite{jensen2016algorithm} and scale-invariant signal-to-distortion ratio (SI-SDR)~\cite{le2019sdr} as evaluation metrics. 
Higher scores mean better
performance for all the metrics.

\subsection{Results}
\label{ssec:results}

\subsubsection{Comparison with competitive baselines}

\begin{table}[t]
\caption{NASE \emph{vs.} other methods. 
``G" and ``D" denote generative and discriminative categories.
* means self-reproduced results.
We select the top-3 open-sourced diffusion SE models as our backbone.}
\centering
\resizebox{0.45\textwidth}{!}{%
\begin{tabular}{p{7.5em}cccc}
\toprule[1.2pt]
System & Category & PESQ & ESTOI & SI-SDR    \\ \midrule[1.2pt]
Unprocessed & - & 1.97 & 0.79 & 8.4 \\ 
\midrule
Conv-TasNet~\cite{luo2019conv} & D & 2.84 & 0.85 & 19.1 \\
GaGNet~\cite{li2022glance} & D & 2.94 & \textbf{0.86} & \textbf{19.9} \\
MetricGAN+~\cite{fu2021metricgan+} & D & \textbf{3.13} & 0.83 & 8.5 \\
\midrule
SEGAN~\cite{pascual2017segan} & G & 2.16 & - & - \\
SE-Flow~\cite{strauss2021flow} & G & 2.28 & - & - \\
RVAE~\cite{bie2022unsupervised} & G & 2.43 & 0.81 & 16.4 \\
CDiffuSE~\cite{lu2022conditional} & G & 2.52 & 0.79 & 12.4 \\
MOSE~\cite{chen2023metric} & G & 2.54 & - & - \\
UNIVERSE*\cite{serra2022universal} & G & 2.91 & 0.84 & 10.1 \\
StoRM~\cite{lemercier2022storm} & G & 2.93 & \textbf{0.88} & 18.8 \\
SGMSE+~\cite{richter2023speech} & G & 2.93 & 0.87 & 17.3 \\
GF-Unified~\cite{shi2023diffusion} & G & 2.97 & 0.87 & 18.3 \\
\midrule
\textbf{NASE} (CDiffuSE) & G & 2.57 & 0.80 & 12.8 \\
\textbf{NASE} (StoRM) & G & 2.98 &\textbf{0.88} & \textbf{18.9} \\
\textbf{NASE} (SGMSE+) & G & \textbf{3.01} & 0.87 & 17.6 \\
\bottomrule[1.2pt]
\end{tabular}}
\label{table1}
\end{table}

Table~\ref{table1} illustrates the comparison of our proposed NASE with competitive baselines, especially the three diffusion SE models that we select as backbones, \textit{i.e.}, CDiffuSE, StoRM and SGMSE+.
Our proposed NASE is shown to be a plug-and-play module and can generalize to various diffusion models for improvement (2.52$\rightarrow$2.57, 2.93$\rightarrow$2.98, 2.93$\rightarrow$3.01), where we have achieved the most 0.08 PESQ improvement on SGMSE+ backbone.
As a result, our NASE has achieved the state-of-the-art among generative SE approaches, though still lagging behind the state-of-the-art discriminative counterparts.
Apart from PESQ metric, our NASE also improves the ESTOI and SI-SDR metrics to some extent.

\subsubsection{Generalization to unseen testing noises}

\begin{table}[t]
\caption{PESQ results on on unseen noise with different SNRs. ``Avg." denotes the average of all SNR levels.}
\resizebox{0.49\textwidth}{!}{\begin{tabular}{lcccccl}
\toprule[1.2pt]
\multirow{2}{*}{System}         &   \multicolumn{5}{c}{Noise level, SNR (dB) =} \\
 & -5 & 0 & 5 & 10 & 15 & Avg. \\
\midrule[1.2pt] 
\multicolumn{7}{c}{\cellcolor[HTML]{EBEBEB}\emph{Noise type: Helicopter}} \\ 
Unprocessed & 1.06 & 1.09 & 1.16 & 1.33 & 1.62 & 1.25 \ \textcolor{teal}{+0\%} \\ 
SGMSE+ & 1.08 & 1.22 & 1.49 & 1.88 & 2.33 & 1.60 \ \textcolor{teal}{+28.0\%} \\ 
\textbf{NASE} (SGMSE+) & \textbf{1.09} & \textbf{1.25} & \textbf{1.57} & \textbf{2.01} & \textbf{2.42} & \textbf{1.67} \ \textcolor{red}{+33.6\%} \\ 
\midrule
\multicolumn{7}{c}{\cellcolor[HTML]{EBEBEB}\emph{Noise type: Baby-cry}} \\ 
Unprocessed & 1.09 & 1.12 & 1.18 & 1.30 & 1.50 & 1.24 \ \textcolor{teal}{+0\%} \\ 
SGMSE+ & 1.21 & 1.44 & 1.85 & 2.34 & 2.83 & 1.93 \ \textcolor{teal}{+55.6\%} \\ 
\textbf{NASE} (SGMSE+) & \textbf{1.24} & \textbf{1.49} & \textbf{1.94} & \textbf{2.43} & \textbf{2.92} & \textbf{2.00} \ \textcolor{red}{+61.3\%} \\ 
\midrule
\multicolumn{7}{c}{\cellcolor[HTML]{EBEBEB}\emph{Noise type: Crowd-party}} \\ 
Unprocessed & 1.13 & 1.14 & 1.21 & 1.34 & 1.58 & 1.28 \ \textcolor{teal}{+0\%} \\ 
SGMSE+~ & 1.26 & 1.58 & 2.02 & 2.42 & 2.83 & 2.02 \ \textcolor{teal}{+57.8\%} \\ 
\textbf{NASE} (SGMSE+) & \textbf{1.28} & \textbf{1.63} & \textbf{2.07} & \textbf{2.49} & \textbf{2.89} & \textbf{2.07} \ \textcolor{red}{+61.7\%} \\ 

\bottomrule[1.2pt]
\end{tabular}}
\label{table2}
\vspace{-0.3cm}
\end{table}

We also evaluate our proposed NASE on three unseen testing noises~\cite{lin2021unsupervised} with a wide range of SNR levels from -5dB to 15dB, where the best SGMSE+ backbone is selected for this study. 
Table~\ref{table2} shows the performance comprison in terms of PESQ metric. 
Despite the outstanding performance on matched VBD test set, we observe that the PESQ result of SEMSE+ baseline on unseen noises dramatically degrades due to noise domain mismatch.
In comparison, our NASE significantly outperforms SGMSE+ in different noise types and SNR levels (1.60$\rightarrow$1.67, 1.93$\rightarrow$2.00, 2.02$\rightarrow$2.07), thanks to the noise-specific information provided by extracted noise conditioner.

In addition, we also find that NASE achieves higher relative PESQ improvement over SGMSE+ on unseen noises, which indicates its effectiveness under varying real-world conditions.
Another interesting fact is that, among the three unseen noises, NASE shows larger relative improvement over SGMSE+ on non-stationary ``Helicopter'' and ``Baby-cry'' noises than stationary ``Crowd-party'' noise, implying the strong robustness of NASE in adverse conditions.

\subsubsection{Effect of audio pre-training in noise classification}

\begin{table}[t]
\caption{Effect of audio pre-training in noise classification module.
``PT'' denotes loading pre-trained BEATs, ``Freeze'' denotes freezing the model parameters of BEATs.}
\vspace{-0.1cm}
\centering
\resizebox{0.45\textwidth}{!}{%
\begin{tabular}{clccccc}
\toprule[1.2pt]
 ID & System & PT & Freeze & PESQ & ESTOI & SI-SDR \\ 
\midrule[1.2pt]
1 &  Unprocessed & - & - & 1.97 & 0.79 & 8.4 \\ 
\midrule
2 & SGMSE+ & - & - & 2.93 & \textbf{0.87} & 17.3 \\
\midrule
3 & \multirow{3}{*}{\textbf{NASE} (SGMSE+)} & \xmark & \xmark & 2.95 & 0.86 & 17.3 \\ 
4 &  & \cmark & \cmark & 2.97 & 0.86 & 17.4 \\ 
5 &  & \cmark & \xmark & \textbf{3.01} & \textbf{0.87} & \textbf{17.6} \\ 
\bottomrule[1.2pt]
\end{tabular}}
\label{table3}
\vspace{-0.1cm}
\end{table}

Table~\ref{table3} illustrates the effect of audio pre-training from BEATs.
First, comparing system 2 and 3, we can observe limited PESQ improvement brought by NASE when without audio pre-training, where the ESTOI and SI-SDR metrics even degrade.
System 4 load the pre-trained BEATs but freeze its parameters during training, which can bring some PESQ improvement (2.95$\rightarrow$2.97).
It indicates that the pre-trained BEATs can produce high-quality but not optimal noise conditioner to improve the reverse process.
In comparison, unfreezing the pre-trained BEATs for training can produce better noise conditioner to guide the reverse denoising process, which thus yields the best SE performance in terms of all three metrics, \textit{i.e.}, system 5.

\vspace{-0.15cm}
\subsubsection{Effect of the weight of noise classification}

\begin{table}[t]
\caption{Effect of the weight of noise classification in multi-task learning.
``Acc.'' denotes the classification accuracy on training data.}
\centering
\resizebox{0.45\textwidth}{!}{%
\begin{tabular}{clccccc}
\toprule[1.2pt]
 ID & System & $\lambda_\text{NC}$ & Acc. (\%) & PESQ & ESTOI & SI-SDR \\ 
\midrule[1.2pt]
6 &  Unprocessed & - & - & 1.97 & 0.79 & 8.4 \\ 
\midrule
7 & SGMSE+ & - & - & 2.93 & 0.87 & 17.3 \\
\midrule
8 & \multirow{5}{*}{\textbf{NASE} (SGMSE+)} & 0 & - & 2.98 & 0.86 & 17.3  \\
9 &  & 0.1 & 71.3 & 3.00 & \textbf{0.88} & 17.4 \\ 
10 &  & 0.3 & 77.4 & \textbf{3.01} & 0.87 & \textbf{17.6} \\ 
11 &  & 0.5 & 81.8 & 2.96 & 0.86 & \textbf{17.6} \\ 
12 &  & 1.0 & 83.6 & 2.92 & 0.86 & 17.2 \\ 
\bottomrule[1.2pt]
\end{tabular}}
\label{table4}
\vspace{-0.2cm}
\end{table}

Table~\ref{table4} analyzes the effect of the weight of noise classification in multi-task learning, which all follow the settings of system 5.
First, system 8 sets $\lambda_\text{NC}$ to 0, which means the Transformer encoder $E$ would be optimized by $\mathcal{L}_\text{diff}$ only.
However, this operation seems insufficient to improve noise conditioner, as compared with system 4 (2.97$\rightarrow$2.98).
On top of that, we start to increase $\lambda_\text{NC}$ to incorporate noise classification into multi-task learning.
System 9 and 10 achieve promising improvements over system 8 in terms of all three metrics, where $\lambda_\text{NC}=0.3$ yields the best SE performance.
Meanwhile, they also perform well in the auxiliary noise classification task, with up to 77.4\% in accuracy.
Further increasing the weight of noise classification task produces higher accuracy up to 83.6\%, but the PESQ performance significantly degrades (3.01$\rightarrow$2.96$\rightarrow$2.92).
This phenomenon indicates that the auxiliary NC task can benefit the diffusion SE with a relatively small weight, by enhancing the conditioner's noise-specificity.
However, when the weight increases, the training of encoder would be dominated by NC task and thus degrade the performance of our targeted SE task~\cite{hu2023gradient}.

\vspace{-0.15cm}
\subsubsection{Effect of different techniques to inject noise conditioner}

\begin{table}[t]
\caption{Effect of different techniques to inject the conditioner, including addition, concatenation and cross-attention fusion.}
\centering
\resizebox{0.45\textwidth}{!}{%
\begin{tabular}{clcccc}
\toprule[1.2pt]
 ID & System & Inject & PESQ & ESTOI & SI-SDR \\ 
\midrule[1.2pt]
13 &  Unprocessed & - & 1.97 & 0.79 & 8.4 \\ 
\midrule
14 & SGMSE+ & - & 2.93 & \textbf{0.87} & 17.3 \\
\midrule
15 & \multirow{3}{*}{\textbf{NASE} (SGMSE+)} & \emph{addition} & \textbf{3.01} & \textbf{0.87} & \textbf{17.6} \\ 
16 &  & \emph{concat} & 2.99 & \textbf{0.87} & 17.5 \\
17 &  & \emph{cross-attn} & 2.96 & 0.86 & 17.5 \\ 
\bottomrule[1.2pt]
\end{tabular}}
\label{table5}
\end{table}

Table~\ref{table5} presents the results of different techniques to inject noise conditioner $E(y)$ into reverse process.
As introduced in Section.~\ref{ssec:noise_cond}, we inject the noise conditioner into $\epsilon_\theta$ by combining it with original inputs, \textit{i.e.}, $x_t$ and $y$.
Here we select three common techniques for feature fusion, \textit{i.e.}, simple addition, feature concatenation and cross-attention fusion.
Our results indicate that all three techniques are effective, where the simple addition yields surprisingly the best performance.
One possible explanation is, in fact the noisy speech $y$ inherently contains noise-related information but not noise-specific, and our extracted noise conditioner exactly serves to complement and highlight the noise-specific information.
Therefore, simple addition or concatenation seems enough to achieve good improvement, while cross-attention may wrongly discard some noise-specific parts and thus leads to sub-optimal performance.

\vspace{-0.1cm}
\subsubsection{Visualization of noise conditioners}

\begin{figure}[t]
\centering
\includegraphics[width=0.43\textwidth]{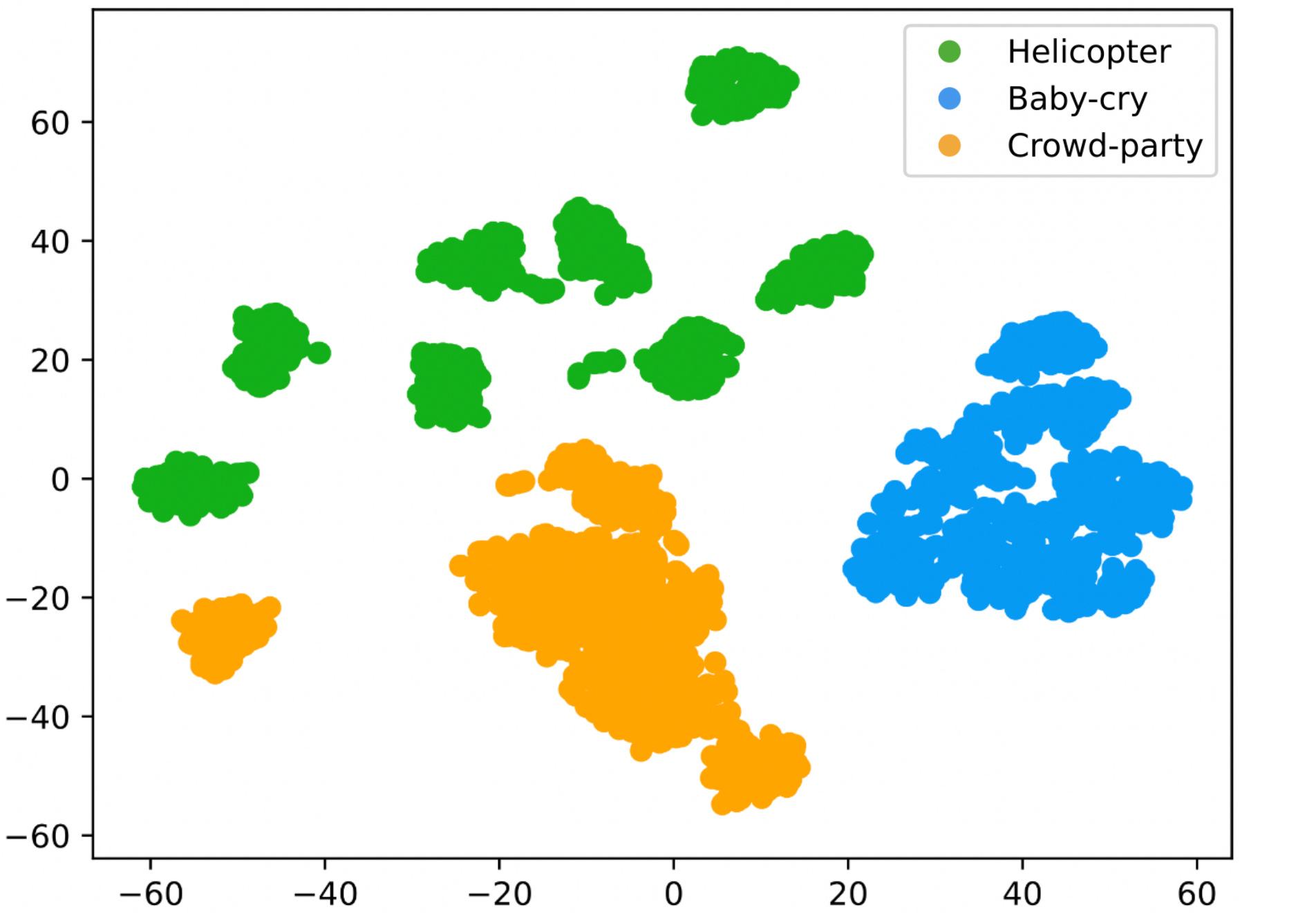}
\caption{The t-SNE visualization of noise conditioners from three unseen noise types, \textit{i.e.}, ``Helicopter'', ``Baby-cry'' and ``Crowd-party''.}
\label{fig2}
\vspace{-0.4cm}
\end{figure}

Fig.~\ref{fig2} visualizes the noise conditioners from three types of unseen noises.
We observe that different noise conditioners are well separated with clear boundaries, indicating their strong noise-specificity.
It guides the reverse process to target at the noise component in $x_t$ for more effective denoising, which is exactly the key to the performance improvement of our NASE.

\section{Conclusion}
\label{sec:conclusion}
In this paper, we propose a noise-aware speech enhancement (NASE) approach that extracts noise-specific information to guide the reverse process of diffusion model.
Specifically, we design a noise classification model and extract its acoustic embedding as a noise conditioner for guiding the reverse process.
Meanwhile, a multi-task learning scheme is devised to jointly optimize SE and NC tasks, aiming to enhance the noise specificity of extracted conditioner.
Our NASE is a plug-and-play module that can generalize to any diffusion SE models.
Experiments verify its effectiveness on multiple diffusion backbones, especially on unseen testing noises.

\bibliographystyle{IEEEtran}
\bibliography{mybib}

\end{document}